\documentclass{aa}  

\usepackage{graphicx}
\usepackage{txfonts}
\usepackage{xcolor}
\usepackage{hyperref}
\hypersetup{colorlinks, linkcolor=blue, citecolor=blue, urlcolor=blue}
\usepackage{amsmath}

\newcommand{\pivec}{\mbox{\boldmath $\pi$}}

\newcommand{\te}{t_{\rm E}}
\newcommand{\thetae}{\theta_{\rm E}}

\newcommand{\pie}{\pi_{\rm E}}
\newcommand{\pien}{\pi_{{\rm E},N}}
\newcommand{\piee}{\pi_{{\rm E},E}}
\newcommand{\dl}{D_{\rm L}}

\definecolor{brown}{rgb}{0.59, 0.29, 0.0}
\definecolor{darkgreen}{rgb}{0.0, 0.42, 0.24}
\definecolor{darkblue}{rgb}{0.01, 0.31, 0.59}
\definecolor{darkblue}{rgb}{0.0, 0.25, 0.42}
\definecolor{blue}{rgb}{0.0,0.0,1.0}
\definecolor{green}{rgb}{0.0,1.0,0.0}

\begin{document}

\title{Four Microlensing Planets with Faint-source Stars Identified in the 2016 and 2017 Season Data}

\author{
     Cheongho~Han\inst{1} 
\and Andrzej~Udalski\inst{2} 
\and Doeon~Kim\inst{1}
\and Youn~Kil~Jung\inst{3} 
\and Yoon-Hyun~Ryu\inst{3} 
\\
(Leading authors)\\
and \\
     Michael~D.~Albrow\inst{5} 
\and Sun-Ju~Chung\inst{3,6}  
\and Andrew~Gould\inst{7,8}
\and Kyu-Ha~Hwang\inst{3} 
\and Chung-Uk~Lee\inst{3} 
\and In-Gu~Shin\inst{3} 
\and Yossi~Shvartzvald\inst{9} 
\and Jennifer~C.~Yee\inst{4} 
\and Weicheng~Zang\inst{10}
\and Sang-Mok~Cha\inst{3,11} 
\and Dong-Jin~Kim\inst{3} 
\and Hyoun-Woo~Kim\inst{3} 
\and Seung-Lee~Kim\inst{3,6} 
\and Dong-Joo~Lee\inst{3} 
\and Yongseok~Lee\inst{3,11} 
\and Byeong-Gon~Park\inst{3,6} 
\and Richard~W.~Pogge\inst{8}
\and Chun-Hwey Kim\inst{12}
\and Woong-Tae Kim\inst{13}
\\
(The KMTNet Collaboration),\\
     Przemek~Mr{\'o}z\inst{2,14} 
\and Micha{\l}~K.~Szyma{\'n}ski\inst{2}
\and Jan~Skowron\inst{2}
\and Radek~Poleski\inst{8} 
\and Igor~Soszy{\'n}ski\inst{2}
\and Pawe{\l}~Pietrukowicz\inst{2}
\and Szymon~Koz{\l}owski\inst{2} 
\and Krzysztof~Ulaczyk\inst{15}
\and Krzysztof~A.~Rybicki\inst{2}
\and Patryk~Iwanek\inst{2}
\and Marcin~Wrona\inst{2}
\\
(The OGLE Collaboration)\\
}

\institute{
     Department of Physics, Chungbuk National University, Cheongju 28644, Republic of Korea  \\ \email{\color{blue} cheongho@astroph.chungbuk.ac.kr}     
\and Warsaw University Observatory, Al.~Ujazdowskie 4, 00-478 Warszawa, Poland                                                                           
\and Korea Astronomy and Space Science Institute, Daejon 34055, Republic of Korea                                                                        
\and Center for Astrophysics $|$ Harvard \& Smithsonian 60 Garden St., Cambridge, MA 02138, USA                                                          
\and University of Canterbury, Department of Physics and Astronomy, Private Bag 4800, Christchurch 8020, New Zealand                                     
\and Korea University of Science and Technology, 217 Gajeong-ro, Yuseong-gu, Daejeon, 34113, Republic of Korea                                           
\and Max Planck Institute for Astronomy, K\"onigstuhl 17, D-69117 Heidelberg, Germany                                                                    
\and Department of Astronomy, The Ohio State University, 140 W. 18th Ave., Columbus, OH 43210, USA                                                       
\and Department of Particle Physics and Astrophysics, Weizmann Institute of Science, Rehovot 76100, Israel                                               
\and Department of Astronomy and Tsinghua Centre for Astrophysics, Tsinghua University, Beijing 100084, China                                            
\and School of Space Research, Kyung Hee University, Yongin, Kyeonggi 17104, Republic of Korea                                                           
\and Department of Astronomy \& Space Science, Chungbuk National University, Cheongju 28644, Republic of Korea                                           
\and Department of Physics \& Astronomy, Seoul National University, Seoul 08826, Republic of Korea                                                       
\and Division of Physics, Mathematics, and Astronomy, California Institute of Technology, Pasadena, CA 91125, USA                                        
\and Department of Physics, University of Warwick, Gibbet Hill Road, Coventry, CV4 7AL, UK                                                               
}
\date{Received ; accepted}

\abstract
{}
{
Microlensing planets occurring on faint source stars can escape detection due to their weak signals. 
Occasionally, detections of such planets are  not reported 
due to the difficulty of extracting high-profile scientific issues on the detected planets.
}
{
For the solid demographic census of microlensing planetary systems based on a complete sample, 
we investigate the microlensing data obtained in the 2016 and 2017 seasons
to search for planetary signals in faint-source lensing events.
From this investigation, we find four unpublished microlensing planets including 
KMT-2016-BLG-2364Lb, KMT-2016-BLG-2397Lb, OGLE-2017-BLG-0604Lb, and OGLE-2017-BLG-1375Lb. 
}
{
We analyze the observed lensing light curves and determine their lensing parameters.  From Bayesian 
analyses conducted with the constraints from the measured parameters, it is found that 
the masses of the hosts and planets are in the ranges $0.50\lesssim M_{\rm host}/M_\odot\lesssim 0.85$ 
and $0.5 \lesssim M_{\rm p}/M_{\rm J}\lesssim 13.2$, respectively, indicating that all planets are 
giant planets around host stars with subsolar masses.  The lenses are located in the distance range 
of $3.8 \lesssim \dl/{\rm kpc}\lesssim 6.4$.  It is found that the lenses of OGLE-2017-BLG-0604 and 
OGLE-2017-BLG-1375 are likely to be in the Galactic disk.
}
{}


\maketitle
%

\section{Introduction}\label{sec:one}

Although the probability for a source star to be gravitationally lensed does not depend on the 
source brightness, the chance to detect microlensing planets decreases as the source becomes 
fainter. This is because the signal-to-noise ratio of the planetary signal in the lensing 
light curve of a faint source event is low due to large photometric uncertainties, and thus, 
if other conditions are the same, the planet detection efficiency of a faint source event is 
lower than that of a bright source event \citep{Jung2014}.  Even if faint source planetary 
events are found despite their lower detection efficiency, they are occasionally left unpublished.  
The major reason for this is that it is difficult to extract high-profile scientific issues on 
the detected planets.  Important scientific issues on microlensing planets are usually found when 
the physical parameters of the planetary lens systems, such as the mass $M$ and distance $\dl$, are 
well constrained.  
For the determinations of these parameters, it is required to simultaneously measure the angular Einstein 
radius, $\thetae$, and the microlens parallax, $\pie$.  The measurement of $\thetae$ requires resolving 
caustic crossings in lensing light curves. For faint source events, the chance to measure $\thetae$ is 
low not only because of the low photometric precision but also because fainter sources tend to be smaller 
stars and thus have a smaller angular source radius, $\theta_*$, and as a result, a shorter duration 
caustic crossing.
i.e., 
\begin{equation}
\Delta t_{\rm cc}={\rho \over \sin \psi} \te;\qquad 
\rho={\theta_* \over \thetae}.
\label{eq1}
\end{equation}
Here $\psi$ denotes the caustic entrance angle of the source star.  The measurement of $\pie$ 
calls for the detection of subtle deviations in the lensing light curve induced by microlens 
parallax effects \citep{Gould1992}, but this measurement is usually difficult for faint source 
events due to the low photometric precision.

Reporting discovered planets is important because, otherwise, the planets would not be included 
in the planet sample to be used for the statistical investigation of the planet properties and 
frequency, e.g., \citet{Gould2010b}, \citet{Sumi2010}, \citet{Suzuki2018}.
As of the time of writing this article, there are 119 microlensing planets in 108 planetary 
systems\footnote{``The Extrasolar Planets Encyclopaedia'' (http://exoplanet.eu/)}.  However, the 
solid characterization of planet properties based on the demographic census of microlensing planets 
requires publishing the discoveries of all microlensing planets including those with faint sources.

In this paper, we report four planetary systems found from the investigation of faint-source 
microlensing events discovered in the 2016 and 2017 seasons.  For the presentation of the work, 
we organize the paper as follows.  In Section~\ref{sec:two}, we state the selection procedure 
of the analyzed planetary lensing events and the data used for the analyses.  In 
Section~\ref{sec:three}, we describe the analysis method commonly applied to the events 
and mention the results from the analyses of the individual events.  In Section~\ref{sec:four}, 
we characterize the source stars of the events by  measuring their color and brightness, and estimate 
$\thetae$ of the events with measured finite-source effects.  In Section~\ref{sec:five}, we estimate 
the physical lens parameters by applying the constraints from the available observables related to 
the parameters.  We summarize the results and conclude in Section~\ref{sec:six}.

\section{Event Selection and Data}\label{sec:two}

Planetary microlensing signals can escape detection in the myriad of data obtained by lensing 
surveys.  Updated microlensing data are available in the public domain so that anomalies of 
various types in lensing light curves can be found for events in progress and trigger decisions 
to conduct followup observations, if necessary, for dense coverage of anomalies, e.g., the OGLE 
Early Warning System \citep{Udalski1994}, MOA Alert System \citep{Bond2001}, and KMTNet Alert 
Finder System: \citep{Kim2018b}.  To make this process successfully work, several modelers 
(V.~Bozza, Y.~Hirao, D.~Bennett, M.~Albrow, Y.~K.~Jung, Y.~H.~Ryu, A. Cassan, and W.~Zang) investigate 
the light curves of lensing events in real time, find anomalous events, and circulate possible 
interpretations of the anomalies to the researchers in the microlensing community.  Despite 
these efforts, some planets may escape detection for various reasons.  One of these reasons is 
the large number of lensing events. During the first generation surveys, e.g., MACHO 
\citep{Alcock1995} and OGLE-I \citep{Udalski1992} surveys, several dozen lensing events were 
annually detected, and thus individual events could be thoroughly investigated.  However, the 
number of event detections has dramatically increased with the enhanced cadence of lensing surveys, 
e.g., the OGLE-IV \citep{Udalski2015}, MOA \citep{Bond2001}, and KMTNet \citep{Kim2016} surveys, 
using globally distributed multiple telescopes equipped with large-format cameras yielding very 
wide field of view, and the current lensing surveys annually detect about 3000 events.  Among 
these events, some planetary lensing events may not be noticed, especially those with weak 
planetary signals hidden in the scatter or noise of the data.

\begin{table}[t]
\small
\centering
\caption{Coordinates of events\label{table:one}}
\begin{tabular*}{\columnwidth}{@{\extracolsep{\fill}}lcc}
\hline\hline
\multicolumn{1}{c}{Event}                           &
\multicolumn{1}{c}{(R.A., decl.)$_{\rm J2000}$}     &
\multicolumn{1}{c}{$(l, b)$}                        \\
\hline
KB-16-2364  &  (17:42:51.76, -27:26:08.02)                           &  $(0^\circ\hskip-2pt.961, 1^\circ\hskip-2pt.30)   $    \\
KB-16-2397  &  (17:44:51.01, -23:12:04.97)                           &  $(4^\circ\hskip-2pt.807, 3^\circ\hskip-2pt.135)  $    \\
OB-17-0604  &  (17:51:34.00, -30:56:47.6)                            &  $(-1^\circ\hskip-2pt.059, -2^\circ\hskip-2pt.140)$    \\
OB-17-1375  &  (17:56:37.17, -30:18:41.1)                            &   $(0^\circ\hskip-2pt.039, -2^\circ\hskip-2pt.757) $   \\
\hline
\end{tabular*}
\end{table}

Considering the possibility of missing planets, we thoroughly investigate the microlensing data of the
OGLE and KMTNet surveys obtained in the 2016 and 2017 seasons, paying special attention to events 
that occurred on faint source stars.  From this investigation, we find one planetary event that was 
not previously known. This event is KMT-2016-BLG-2364.  We also investigate events with known planets, 
for which the findings have not been published or there is no plan to publish on the individual event 
basis.  There exist three such events including KMT-2016-BLG-2397, OGLE-2017-BLG-0604, and OGLE-2017-BLG-1375.  
In this work, we present detailed analysis on these four planetary events.

The analyzed planetary lensing events are located toward the Galactic bulge field. The positions,
both in equatorial and Galactic coordinates, of the individual events are listed in Table~\ref{table:one}. 
All these events occurred in the fields that were commonly observed by both surveys.  The event
OGLE-2017-BLG-1375/KMT-2017-BLG-0078, hereafter we refer to as OGLE-2017-BLG-1375 as a representative 
name of the event according to the chronological order of the event discovery, was found by both surveys.  
However, the other events were detected by only a single survey and escaped detection by the other survey.  
As a result, the two events KMT-2016-BLG-2364 and KMT-2016-BLG-2397 were found only by the KMTNet survey, 
and the event OGLE-2017-BLG-0604 was found solely by the OGLE survey.  Although these events are detected 
by a single survey, we use data from both surveys in our analysis by 
conducting post-event photometry
of the events.

Observations of the events by the OGLE survey were conducted using the 1.3~m telescope of the Las 
Campanas Observatory in Chile. The telescope is equipped with a camera yielding a 1.4~deg$^2$ field 
of view. The KMTNet survey used three identical 1.6~m telescopes located at the Siding Spring 
Observatory in Australia (KMTA), Cerro Tololo Interamerican Observatory in Chile (KMTC), and the 
South African Astronomical Observatory in South Africa (KMTS). Hereafter, we refer to the individual 
KMTNet telescopes as KMTA, KMTC, and KMTS, respectively. Each KMTNet telescope is equipped with a 
camera providing a 4 deg$^2$ field of view. For both surveys, observations were carried out mainly in 
the $I$ band, and $V$ band observations were conducted for a subset of images to measure the source 
color.  We give detailed description about the procedure of the source color measurement in 
Section~\ref{sec:four}.

\begin{figure}
\includegraphics[width=\columnwidth]{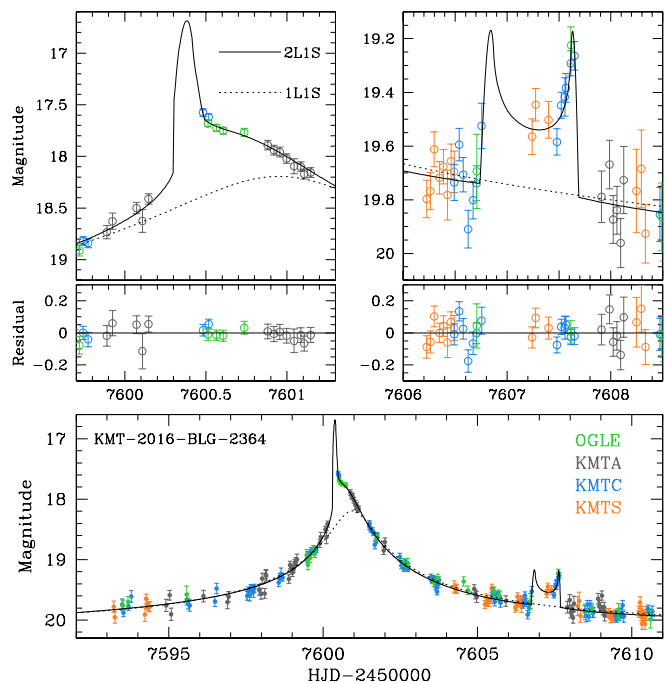}
\caption{
Light curve of KMT-2016-BLG-2364. The dotted and solid curves superposed on the data points are 
the lL1S and 2L1S models, respectively.  The lensing parameters of the 2L1S model are presented in 
Table~\ref{table:two}. The upper panels show the enlarged views of the regions around 
${\rm HJD}^\prime\sim 7601$ (left panel) and $\sim 7607$ (right panel) when the planet-induced 
anomalies occur.  The lens system configuration of the 2L1S solution is presented in 
Figure~\ref{fig:two}.
}
\label{fig:one}
\end{figure}

\begin{table}[t]
\small
\caption{Lensing parameters of KMT-2016-BLG-2364\label{table:two}}
\begin{tabular*}{\columnwidth}{@{\extracolsep{\fill}}lc}
\hline\hline
\multicolumn{1}{c}{Parameter}     &
\multicolumn{1}{c}{Value}         \\
\hline
$t_0$ (${\rm HJD}^\prime$)    &  $7600.803 \pm 0.022$    \\
$u_0$                         &  $0.028 \pm 0.002   $    \\
$\te$ (days)                  &  $20.33 \pm 0.96    $    \\
$s$                           &  $1.169 \pm 0.008   $    \\
$q$ ($10^{-3}$)               &  $7.56 \pm 0.72     $    \\
$\alpha$ (rad)                &  $3.026 \pm 0.010   $    \\
$\rho$                        &  $< 0.003           $    \\
$f_s$ / $I_s$                 &  0.028 / 21.87           \\
\hline
\end{tabular*}
\tablefoot{ 
${\rm HJD}^\prime = {\rm HJD}- 2450000$.  
The source flux $f_s$ is on an $I=18$ scale, i.e., $f_s = 10^{-0.4(I_s-18)}$.
}
\end{table}

Photometry of the events was carried out using the software pipelines developed by the individual 
survey groups: \citet{Udalski2003} using the DIA technique \citet{Wozniak2000} for the OGLE survey 
and \citet{Albrow2009} for the KMTNet survey. Both pipelines are based on the 
difference imaging method \citep{Tomaney1996, Alard1998}, that is optimized for dense field photometry.  
For a subset of the KMTC $I$- and $V$-band images, we conduct additional photometry using the pyDIA 
software \citep{Albrow2017} for the source color measurement.  For the data used in the analysis, we 
readjust the errorbars of the data  so that the cumulative $\chi^2$ distribution with respect to the 
lensing magnification becomes linear and $\chi^2$ per degree of freedom for each data 
set becomes unity \citep{Yee2012}.

\section{Analyses}\label{sec:three}

Modeling each lensing event is carried out by searching for the parameters that 
best explain the observed light curve.  The light curves of the analyzed events share the 
common characteristics that the events are produced by binary lens and a single source (2L1S) 
with discontinuous caustic-involved features.  The light curves of such 2L1S events are described 
by seven lensing parameters. The first three of these parameters are those of a single-lens 
single-source (1L1S) event, including ($t_0$, $u_0$, $\te$), which denote the time of the closest 
lens approach to the source, the impact parameter of the lens-source encounter (in units of $\thetae$), 
and the event timescale, respectively.  The next three parameters are related to the binarity of the 
lens, and these parameters include $(s, q, \alpha)$, which indicate the projected binary separation 
(in units of $\thetae$), the mass ratio between the lens components, $M_1$ and $M_2$, and the incidence 
angle of the source trajectory as measured from the $M_1$--$M_2$ axis, respectively. The last parameter 
is the normalized source radius $\rho$, which is included in modeling because the light curves of all 
analyzed events exhibit discontinuous features that are likely to be involved with source stars' caustic 
crossings, during which the lensing light curve is affected by finite-source effects.

\begin{figure}
\includegraphics[width=\columnwidth]{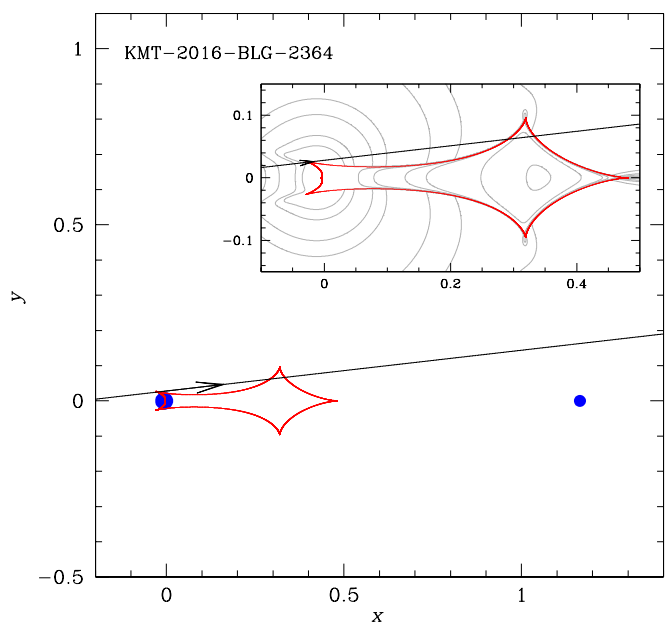}
\caption{
Configuration of the KMT-2016-BLG-2364 lens system.  The line with an arrow denotes the source 
trajectory with respect to the lens components that are marked by blue filled dots.  The cuspy 
closed figure drawn in red color represents the caustic.  The inset shows the enlargement of the 
central magnification region. The grey curves around the caustic represent equi-magnification 
contours.  Lengths are normalized to the angular Einstein radius corresponding to the total 
lens mass.
}
\label{fig:two}
\end{figure}

The procedure of modeling commonly applied to the analysis is as follows.  In the first-round modeling, 
we search for the binary lensing parameters $s$ and $q$ using a grid-search approach, while the other 
lensing parameters are found using a downhill method, which is based on the Markov Chain Monte Carlo 
(MCMC) algorithm.  From the $\Delta\chi^2$ map in the $s$--$q$ plane constructed from this round of 
modeling, we identify local minima. In the second-round modeling, we inspect the individual 
local minima and refine the lensing parameters by releasing all parameters, including $s$ and $q$, 
as free parameters.  We then compare the $\chi^2$ values of the local solutions not only to find a 
global solution but also to find degenerate solutions, if they exist.  In the following subsections, 
we present details of the analysis applied to the individual events and the results from the analysis.

\begin{figure}
\includegraphics[width=\columnwidth]{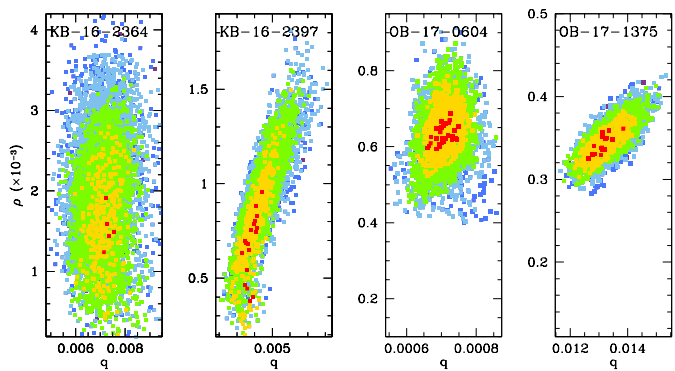}
\caption{
$\Delta\chi^2$ distributions of points in the MCMC chain on the $q$--$\rho$ plane
for the individual lensing events.  The colors of the data points indicate regions with 
$\leq 1\sigma$ (red), 
$\leq 2\sigma$ (yellow), 
$\leq 3\sigma$ (green), 
$\leq 4\sigma$ (cyan), and 
$\leq 5\sigma$ (blue). 
}
\label{fig:three}
\end{figure}

\subsection{KMT-2016-BLG-2364}\label{sec:three-one}
  
KMT-2016-BLG-2364 is a lensing event for which the existence of a very low-mass companion to the 
lens was not known during the season of the event and found from the post-season inspection of the 
2016 season data \citep{Kim2018a}.  Figure~\ref{fig:one} shows the lensing light curve of the event. 
The dotted curve superposed on the data points represents the 1L1S model. The planet-induced anomalies 
appear in the two very localized regions around ${\rm HJD}^\prime\equiv {\rm HJD}-2450000\sim 7600$ 
and $\sim 7607$, for which the zoomed-in views are shown in the upper left and right panels, respectively.  
It was difficult to notice the anomalies at a casual glance of the light curve constructed with the online 
KMTNet data, because the former anomaly (at ${\rm HJD}^\prime\sim 7600$)  was covered by only two data 
points and the latter anomaly (at ${\rm HJD}^\prime\sim 7607$) was hidden in the scatter and noise of the 
online data processed by the automated photometry pipeline.  We inspect the event to check whether 
the two points at ${\rm HJD}^\prime=7600.72$ and 7600.85 are real by first rereducing the data for 
optimal photometry, and then conducting a 2L1S modeling.  This procedure not only confirms the reality 
of the former anomaly but also enabled us to unexpectedly find the latter anomaly.  The anomalous 
features are additionally confirmed with the addition of the OGLE data processed after the analysis 
based on the KMTNet data.

From modeling the light curve, we find that the anomalies are produced by a planetary companion to the 
primary lens with $(s,q)\sim (1.17, 7.6\times 10^{-3})$. The model curve, the solid curve plotted 
over the data points, of the solution is presented in Figure~\ref{fig:one}, and the lensing parameters 
of the solution are listed in Table~\ref{table:two}. We find that the presented solution is unique 
without any degeneracy due to the special lens system configuration (see below) producing the anomalies 
at two remotely separated regions.  Also listed in Table~\ref{table:two} are the flux values and 
magnitudes of the source at the baseline, $f_s$ and $I_s$.  The flux is on an $I=18$ scale, i.e., 
$f = 10^{-0.4(I-18)}$.  It is found that the source of the event is very faint with an apparent 
magnitude of $I_s\sim 21.9$.

The configuration of the lens system is shown in Figure~\ref{fig:two}, in which the trajectory of 
the source (line with an arrow) relative to the positions of the lens components (marked by blue dots) 
and the resulting lensing caustic (red cuspy closed figure) are presented. The lens system induces a 
single large caustic with 6 cusps, i.e., resonant caustic, due to the closeness of the binary separation 
to unity.  According to the solution, the anomalies at ${\rm HJD}^\prime\sim 7600$ and $\sim 7607$ were 
produced when the source passed the upper left and right tips of the caustic, respectively.  We 
check the feasibility of measuring the microlens parallax $\pie$ by conducting additional modeling. 
This modeling requires including the extra parameters $\pien$ and $\piee$, which represent the north 
and east components of $\pivec_{\rm E}$, respectively.  From this, it is found that $\pie$ cannot be 
securely measured, 
mostly due to the relatively large uncertainties of the data caused by the faintness of the source together 
with the relatively short event timescale, $\te \sim21$~days.  Although both anomalies are captured, the 
caustic crossings are poorly resolved, and this makes it difficult to constrain the normalized source 
radius $\rho$.  This is shown in the $\Delta\chi^2$ distribution of points in the MCMC chain on the 
$q$--$\rho$ plane presented in Figure~\ref{fig:three}.  It is found that the observed light curve is 
consistent with a point-source model, although the upper limit is constrained to be $\rho\lesssim 0.003$ 
as measured at the $3\sigma$ level.

\begin{figure}
\includegraphics[width=\columnwidth]{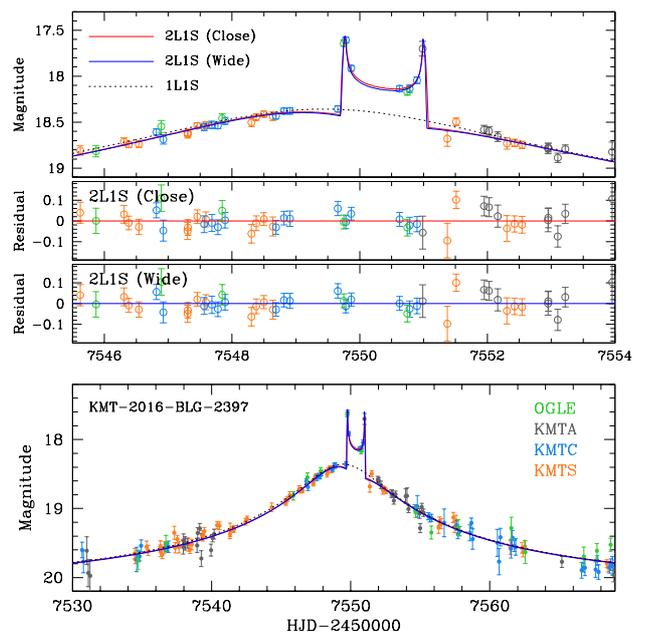}
\caption{
Lensing light curve of KMT-2016-BLG-2397.  The dotted and solid curves are the 1L1S and 2L1S models, 
respectively.  The enlarged view of the anomaly region is shown in the top panel.  The second 
and third panels show the residuals from the ``close'' and ``wide'' 2L1S solutions, for which
the corresponding lensing parameters are presented in Table~\ref{table:three}, and the lens system 
configurations are shown in Fig.~\ref{fig:five}.
}
\label{fig:four}
\end{figure}

\subsection{KMT-2016-BLG-2397}\label{sec:three-two}

The event KMT-2016-BLG-2397 was found by the KMTNet survey, and the OGLE data were recovered from 
the post-event photometry for the lensing source identified by the KMTNet survey.  The anomaly, 
which occurred at ${\rm HJD}^\prime\sim 7550.4$ near the peak, lasted for $\sim 1.5$~days.  The 
planetary origin of the anomaly was known by several modelers of the KMTNet group from the analyses 
of the online data conducted during the 2016 season, but no extended analysis based on optimized 
photometric data has been presented until this work.

\begin{figure}
\includegraphics[width=\columnwidth]{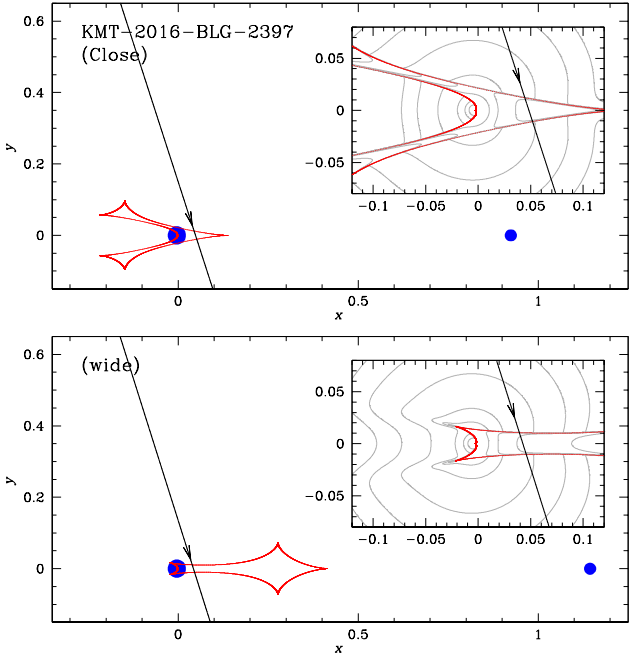}
\caption{
Configuration of the lens system for the lensing event KMT-2016-BLG-2397.  The upper and lower 
panels are those of the close ($s<1.0$) and wide ($s>1.0$) solutions, respectively. Notations 
are same as those in Fig.~\ref{fig:three}.
}
\label{fig:five}
\end{figure}

\begin{table}[thb]
\small
\caption{Lensing parameters of KMT-2016-BLG-2397\label{table:three}}
\begin{tabular*}{\columnwidth}{@{\extracolsep{\fill}}lcc}
\hline\hline
\multicolumn{1}{c}{Parameter}     &
\multicolumn{2}{c}{Value}         \\
\multicolumn{1}{c}{}              &
\multicolumn{1}{c}{Close}         &
\multicolumn{1}{c}{Wide}          \\
\hline
$\chi^2$                      &  653.1                  &  652.6                \\         
$t_0$ (${\rm HJD}^\prime$)    &  $7549.641 \pm 0.047$   &  $7549.651 \pm 0.050$ \\
$u_0$                         &  $0.046 \pm 0.006   $   &  $0.041 \pm 0.006   $ \\
$\te$ (days)                  &  $52.74 \pm 6.53    $   &  $58.32 \pm 8.10    $ \\
$s$                           &  $0.927 \pm 0.012   $   &  $1.148 \pm 0.015   $ \\
$q$ ($10^{-3}$)               &  $3.72 \pm 0.78     $   &  $3.95 \pm 0.92     $ \\
$\alpha$ (rad)                &  $4.402 \pm 0.019   $   &  $4.409 \pm 0.020   $ \\
$\rho$ ($10^{-3}$)            &  $<0.0015           $   &  $\leftarrow        $ \\
$f_s$ / $I_s$                 &  0.032 / 21.74          &  $\leftarrow        $ \\
\hline
\end{tabular*}
\end{table}

The light curve of KMT-2016-BLG-2397 is shown in Figure~\ref{fig:four}, in which the enlarged view 
around the anomaly is presented in the upper panel.  The combination of the six KMTC data points 
plus two OGLE points in the anomaly region display a ``U''-shape pattern, which is a characteristic 
pattern appearing during the passage of a source inside a caustic, indicating that the anomaly is 
produced by caustic crossings.  Modeling the light curve yields two degenerate solutions resulting 
from the close/wide degeneracy \citep{Griest1998, Dominik1999}.  The binary parameters are 
$(s,q)\sim (0.93, 3.72\times 10^{-3})$ and $\sim (1.15, 3.95\times 10^{-3})$ for the close ($s<1.0$) 
and wide ($s>1.0$) solutions, respectively, indicating that the anomaly is generated by a planetary 
companion to the primary lens located near the Einstein ring of the primary.  The degeneracy between 
the close and wide solutions is very severe, and the $\chi^2$ difference between the two degenerate 
models is merely $\Delta\chi^2=0.5$.  The lensing parameters of the two solutions are listed in 
Table~\ref{table:three} together with $\chi^2$ values of the fits.

The lens system configuration of the event is displayed in Figure~\ref{fig:five}.  The upper and 
lower panels show the configurations for the close and wide solutions, respectively, and the inset 
in each panel shows the enlarged view of the central magnification region.  As in the case of 
KMT-2016-BLG-2364, the lens system forms a single large resonant caustic because the binary separation 
is similar to $\thetae$, i.e., $s\sim 1$. For both close and wide solutions, the source trajectory 
passes the planet side of the caustic, producing an anomaly characterized by two spikes and U-shape 
trough region between the spikes.

The light curve of KMT-2016-BLG-2397 shares many characteristics in common with that of KMT-2016-BLG-2364.  
First, the source star of the event is very faint with a baseline magnitude of $I\sim 21.7$.  Second, 
the caustic crossings are not well resolved, and this makes it difficult to constrain the normalized 
source radius $\rho$, although the upper limit is set to be $\rho\lesssim 0.0015$.  See the $\Delta\chi^2$ 
distribution on the $q$--$\rho$ plane presented in Figure~\ref{fig:three}.  Third, due to the substantial 
uncertainty of the photometric data caused by the faintness of the source, the microlens parallax $\pie$ 
cannot be securely determined despite the relatively long timescale, which is $\te\sim 53$~days for the 
close solution and $\te\sim 58$ days for the wide solution, of the event.

\begin{figure}
\includegraphics[width=\columnwidth]{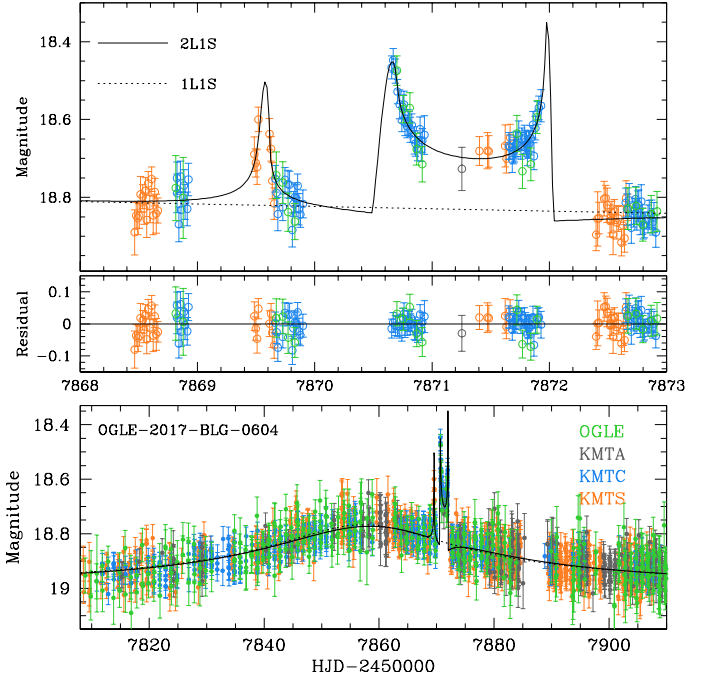}
\caption{
Light curve of OGLE-2017-BLG-0604.  The enlarged view around the anomaly is presented in the upper 
panel.  Notations are same as those in Fig.~\ref{fig:one}.  The lensing parameters of the solutions 
are presented in Table~\ref{table:four}, and the corresponding lens system configuration is shown in 
Fig.~\ref{fig:seven}.  
}
\label{fig:six}
\end{figure}

\subsection{OGLE-2017-BLG-0604}\label{sec:three-three}

The lensing event OGLE-2017-BLG-0604 was found by the OGLE survey.  The source star was very 
faint, with a baseline magnitude of $I\sim 21.6$.  Together with the low magnification, with 
a peak magnification $A_{\rm peak}\sim 3.3$, the event was not detected by the event-finder 
system of the KMTNet survey \citep{Kim2018a}.  The existence of a possible short-term anomaly 
in the OGLE light curve was noticed by KMTNet modelers.  Realizing that the event is located 
in the KMTNet field, we conduct photometry for the KMTNet images at the location of the source 
using the finding chart provided by the OGLE survey and recover the KMTNet light curve of the 
event.  It is found that the additional KMTNet data are crucial in finding a unique solution 
of the event.  See below for more detailed discussions.

\begin{figure}
\includegraphics[width=\columnwidth]{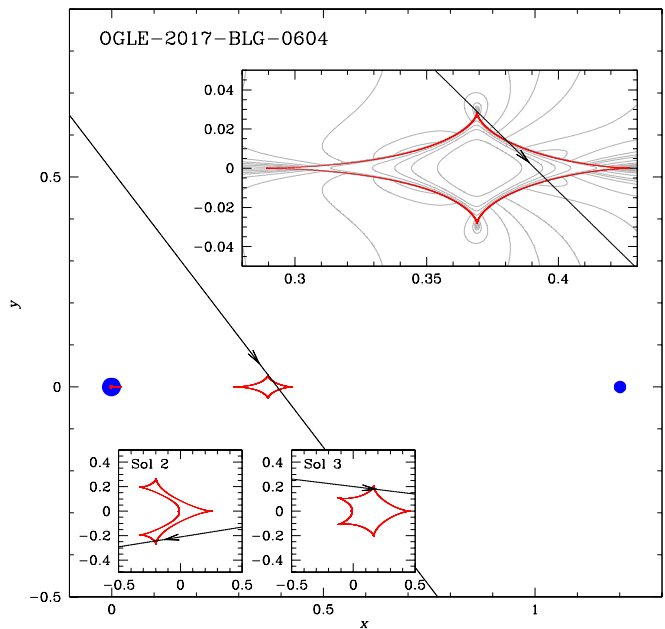}
\caption{
Configuration of the OGLE-2017-BLG-0604 lens system.  The upper right inset shows the close-up view of 
the planetary caustic and the source trajectory.  The two lower left insets show the configurations of 
the two degenerate solutions obtained from modeling with only the OGLE data, which do not cover the first 
part of the anomaly at ${\rm HJD}^\prime\sim 7869.6$.  These degenerate solutions cannot explain the 
first part of the anomaly, although they almost equally well describe the second part of the anomaly 
as the presented solution.
}
\label{fig:seven}
\end{figure}

\begin{table}[htb]
\small
\caption{Lensing parameters of OGLE-2017-BLG-0604\label{table:four}}
\begin{tabular*}{\columnwidth}{@{\extracolsep{\fill}}lc}
\hline\hline
\multicolumn{1}{c}{Parameter}     &
\multicolumn{1}{c}{Value}         \\
\hline
$t_0$ (${\rm HJD}^\prime$)    &  $7858.520 \pm 0.217$   \\
$u_0$                         &  $0.312 \pm 0.007   $   \\
$\te$ (days)                  &  $55.21 \pm 1.13    $   \\
$s$                           &  $1.201 \pm 0.004   $   \\
$q$ ($10^{-3}$)               &  $0.70 \pm 0.04     $   \\
$\alpha$ (rad)                &  $4.064 \pm 0.015   $   \\
$\rho$ ($10^{-3}$)            &  $0.60 \pm 0.09     $   \\
$f_s$ / $I_s$                 &  0.037 / 21.58          \\
\hline
\end{tabular*}
\end{table}

The light curve of OGLE-2017-BLG-0604 is displayed in Figure~\ref{fig:six}. The upper panel shows the 
zoomed-in view of the anomaly region during $7868\lesssim {\rm HJD}^\prime \lesssim 7873$. The 
anomaly is composed of two parts: the brief bump centered at ${\rm HJD}^\prime\sim 7869.6$ (covered 
mainly by the KMTS data set) and the caustic-crossing features between 
$7870.6\lesssim  {\rm HJD}^\prime \lesssim 7872.0$ (covered by all data sets). The first part of 
the anomaly is not obvious in the OGLE data, and the modeling based on only the OGLE data yields 
several possible solutions caused by accidental degeneracies.  However, modeling with the use of 
the additional KMTNet data, clearly showing the first anomaly, yields a unique solution, excluding 
the other solutions found from the modeling  without the KMTNet data.  This indicates that the 
data covering the first part of the anomaly are crucial for the accurate characterization of the 
lens system.

The lensing parameters of the best-fit solution are listed in Table~\ref{table:four}, and the lens 
system configuration corresponding to the solution is shown in Figure~\ref{fig:seven}. The estimated 
binary parameters are $(s,q)\sim (1.20, 0.70\times 10^{-3})$, indicating that the anomalies are 
produced by a planetary companion to the lens.  The planet induces two sets of caustics, in which 
one is located near the location of the primary (central caustic), and the other is located away 
from the host (planetary caustic) toward of direction of the planet with a separation of 
$s-1/s\sim 0.37$ \citep{Griest1998, Han2006}. The anomaly in the lensing light curve was produced 
by the source star's approach and crossings over the planetary caustic.  Before the caustic crossings, 
which produced the major caustic-crossing feature in the lensing light curve, the source approached 
the upper cusp of the caustic, and this produced the first part of the anomaly at 
${\rm HJD}^\prime\sim 7869.6$.  See the enlarged view of the planetary caustic and the source 
trajectory shown in the inset of the upper panel.  The lower two insets show the configurations of 
the lens systems for the two degenerate solutions, with $(s,q)\sim (0.90, 0.027)$ and 
$\sim (1.08, 0.028)$, obtained from the modeling not using the KMTNet data.  These solutions almost 
equally well describe the second part of the anomaly as the presented solution, but cannot explain 
the first part of the anomaly.  The coverage of the first part of the anomaly also enables us to 
measure the normalized source radius of $\rho\sim 0.60\times 10^{-3}$.  See the $\Delta\chi^2$ 
distribution of MCMC points on the $q$--$\rho$ plane shown in Figure~\ref{fig:three}.  However, 
the microlens parallax cannot be determined due to the substantial photometric uncertainty of the 
data caused by the faintness of the source star.

\begin{figure}
\includegraphics[width=\columnwidth]{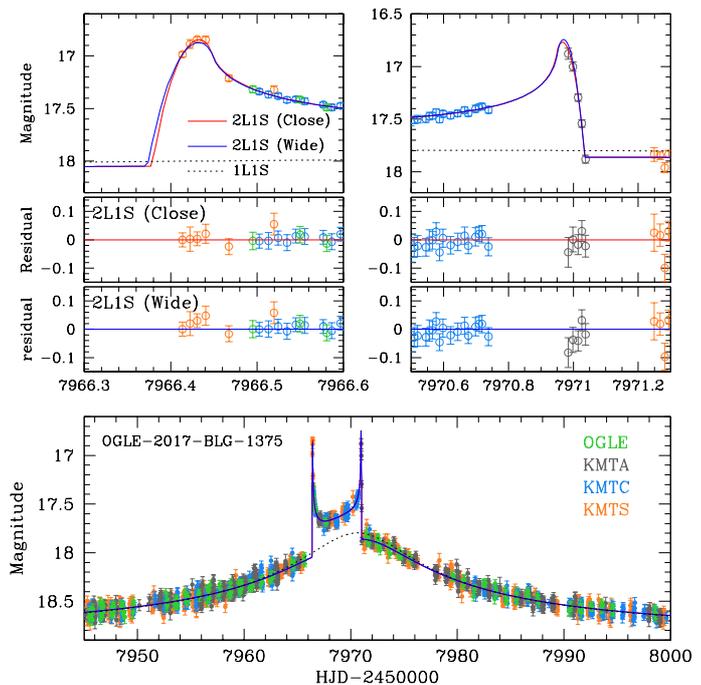}
\caption{
Light curve of OGLE-2017-BLG-1375. The upper left and right panels show the close-up views of the 
caustic entrance and exit of the source, respectively. For each of the upper panels, we present 
two sets of residuals from the close and wide binary solutions.
The lensing parameters of the solutions are presented in Table~\ref{table:five},
and the lens system configuration is shown in Fig.~\ref{fig:nine}.
}
\label{fig:eight}
\end{figure}

\begin{table*}[htb]
\small
\caption{Lensing parameters of OGLE-2017-BLG-1375\label{table:five}}
\begin{tabular}{l|c|c|c|c|c|c}
\hline\hline
\multicolumn{1}{c|}{Parameter}                   &
\multicolumn{3}{c|}{Close}              &
\multicolumn{3}{c}{Wide}               \\  
\multicolumn{1}{c|}{}                   &
\multicolumn{1}{c|}{Standard}           &
\multicolumn{2}{c|}{Orbit+Parallax}     &
\multicolumn{1}{c|}{Standard}           &
\multicolumn{2}{c}{Orbit+Parallax}     \\
\multicolumn{1}{c|}{}                   &
\multicolumn{1}{c|}{}                   &
\multicolumn{1}{c|}{$u_0>0$}                   &
\multicolumn{1}{c|}{$u_0<0$}                   &
\multicolumn{1}{c|}{}                   &
\multicolumn{1}{c|}{$u_0>0$}                   &
\multicolumn{1}{c}{$u_0<0$}                   \\
\hline
$\chi^2$                      &  $5637.0             $  &   $ 5629.4              $   &   $ 5629.3              $  &   $5644.3             $  &  $5636.9            $  &    $5637.1           $ \\
$t_0$ (${\rm HJD}^\prime$)    &  $7969.909 \pm 0.020 $  &   $ 7969.866 \pm 0.026  $   &   $ 7969.858 \pm 0.023  $  &   $7969.919 \pm 0.020 $  &  $7969.645\pm 0.028 $  &    $ 7969.630\pm 0.031 $ \\
$u_0$                         &  $0.035 \pm 0.002    $  &   $ 0.034 \pm 0.001     $   &   $-0.035 \pm 0.001     $  &   $0.035 \pm 0.002    $  &  $0.028 \pm 0.002   $  &    $-0.028 \pm 0.002   $ \\
$\te$ (days)                  &  $101.36 \pm 3.66    $  &   $ 104.27 \pm 3.45     $   &   $ 101.06 \pm 1.55     $  &   $101.81 \pm 3.82    $  &  $103.03 \pm 7.93   $  &    $ 101.98 \pm 5.37   $ \\
$s$                           &  $0.837 \pm 0.004    $  &   $ 0.840 \pm 0.004     $   &   $ 0.839 \pm 0.003     $  &   $1.274 \pm 0.005    $  &  $1.269 \pm 0.005   $  &    $ 1.268 \pm 0.005   $ \\
$q$ ($10^{-3}$)               &  $13.46 \pm 0.49     $  &   $ 12.76 \pm 0.54      $   &   $ 13.32 \pm 0.38      $  &   $15.30 \pm 0.63     $  &  $14.81 \pm 1.15    $  &    $ 14.88 \pm 0.84    $ \\
$\alpha$ (rad)                &  $5.100 \pm 0.005    $  &   $ 5.102 \pm 0.009     $   &   $-5.102 \pm 0.007     $  &   $5.071 \pm 0.005    $  &  $5.068 \pm 0.006   $  &    $-5.069 \pm 0.005   $ \\
$\rho$ ($10^{-3}$)            &  $0.34 \pm 0.02      $  &   $ 0.33 \pm0.02        $   &   $ 0.34 \pm0.02        $  &   $0.36 \pm 0.02      $  &  $0.36 \pm 0.03     $  &    $ 0.36 \pm 0.03     $ \\
$\pi_{{\rm E},N}$             &   --                    &   $  0.069 \pm 0.114$       &   $  0.072 \pm 0.123$      &   --                     &  $  0.134 \pm 0.112 $  &    $   0.004 \pm 0.121 $ \\
$\pi_{{\rm E},E}$             &   --                    &   $ -0.068 \pm 0.025$       &   $ -0.052 \pm 0.018$      &   --                     &  $ -0.064 \pm 0.024 $  &    $  -0.042 \pm 0.021 $ \\
$f_s$ / $I_s$                 &   0.036 / 21.62         &   $\leftarrow$              &   $\leftarrow$             &   $\leftarrow$           &  $\leftarrow$          &    $\leftarrow$        \\
\hline
\end{tabular}
\tablefoot{ 
The notation ``$\leftarrow$'' denotes that the value is same as that in the left side column, and
``--'' indicates that the parameter is not measured or is poorly constrained.
}
\end{table*}

\begin{figure}
\includegraphics[width=\columnwidth]{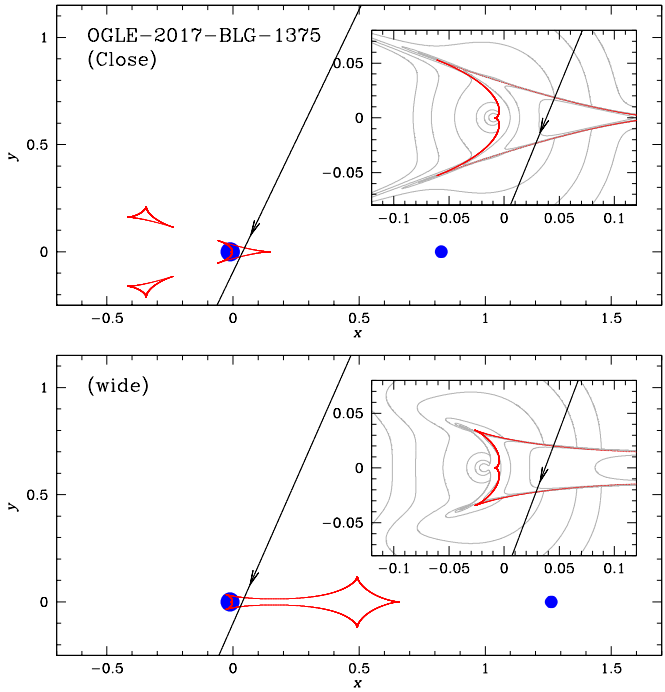}
\caption{
Lens configuration of the event OGLE-2017-BLG-1375. The upper and lower panels show the configurations 
for the close and wide solutions, respectively. Notations are same as those in Fig.~\ref{fig:two}.
}
\label{fig:nine}
\end{figure}

\subsection{OGLE-2017-BLG-1375}\label{sec:three-four}

The event OGLE-2017-BLG-1375 was first found by the Early Warning System \citep{Udalski1994} 
of the OGLE survey on 2017-07-20, ${\rm HJD}^\prime\sim 7954$. Later, the event was 
independently found by the KMTNet Event Finder System \citep{Kim2018a} and it was designated 
as KMT-2017-BLG-0078.  The source is a faint star with an $I$-band magnitude of $I_s\sim 21.6$.

The light curve of the lensing event is displayed in Figure~\ref{fig:eight}. It shows an obvious 
caustic-crossing feature, in which the spikes at ${\rm HJD}^\prime\sim 7966.43$ and $\sim 7970.95$ 
are caused by the caustic entrance and exit of the source star, respectively.  The individual 
caustic crossings were covered by the KMTS (for the caustic entrance) and KMTA (for the caustic 
exit) data sets, respectively.  See the upper panels showing the enlarged views of the 
caustic-crossing parts of the light curve.  The obvious anomaly feature led to real-time modeling 
of the event based on the online data at the time of the anomaly by several KMTNet modelers, but 
no result from a detailed analysis has been reported before this work.

Analysis using the data obtained from optimized photometry indicates that the event is produced 
by a binary lens with a low-mass companion. Interpreting the event is subject to a close/wide 
degeneracy, and the estimated binary parameters are $(s, q)\sim (0.84, 0.013)$ for the close 
solution and $(s, q)\sim (1.27, 0.015)$ for the wide solution. The lensing parameters for both 
solutions based on the rectilinear (linear motion without acceleration) relative lens-source 
motion (``standard solution'') are listed in Table~\ref{table:five}.  Because the caustic 
crossings are resolved, the normalized source radius, $\rho\sim (0.33-0.34)\times 10^{-3}$, is 
well constrained, as shown in the $\Delta\chi^2$ distribution of MCMC points in Figure~\ref{fig:three}.  
Figure~\ref{fig:nine} shows the lens system configurations for the close (upper panel) and wide (lower 
panel) binary solutions. The inset for each configuration shows the close-up view of the central 
magnification region, through which the source passed. According to both solutions, the anomaly 
was produced by the source passage over the planet-side central caustic.

\begin{figure}
\includegraphics[width=\columnwidth]{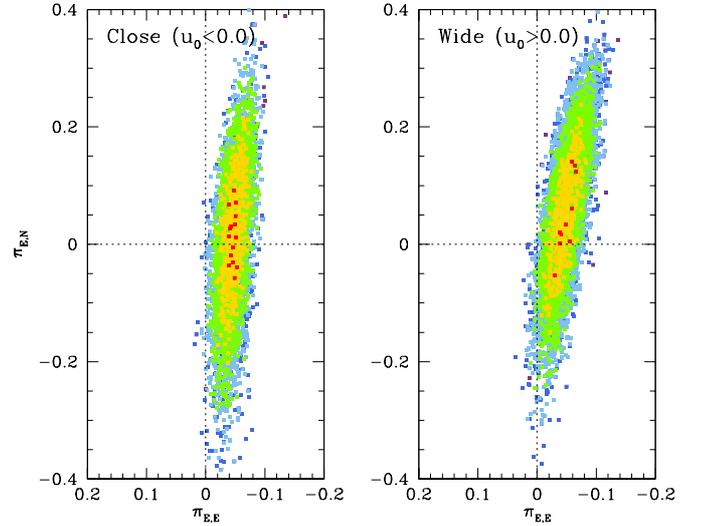}
\caption{
Distribution of points in the MCMC chain on the $\piee$--$\pien$ parameter plane for 
OGLE-2017-BLG-1375.  Left and right panels are for the close (with $u_0<0$) and wide (with 
$u_0>0$) binary solutions, respectively.  The color coding is set to denote points with 
$<1\sigma$ (red), $<2\sigma$ (yellow), $<3\sigma$ (green), $<4\sigma$ (cyan), and $<5\sigma$ 
(blue).  The dotted cross hair represents the lines with $(\piee, \pien)=(0.0, 0.0)$.
}
\label{fig:ten}
\end{figure}

\begin{figure*}
\centering
\includegraphics[width=15.0cm]{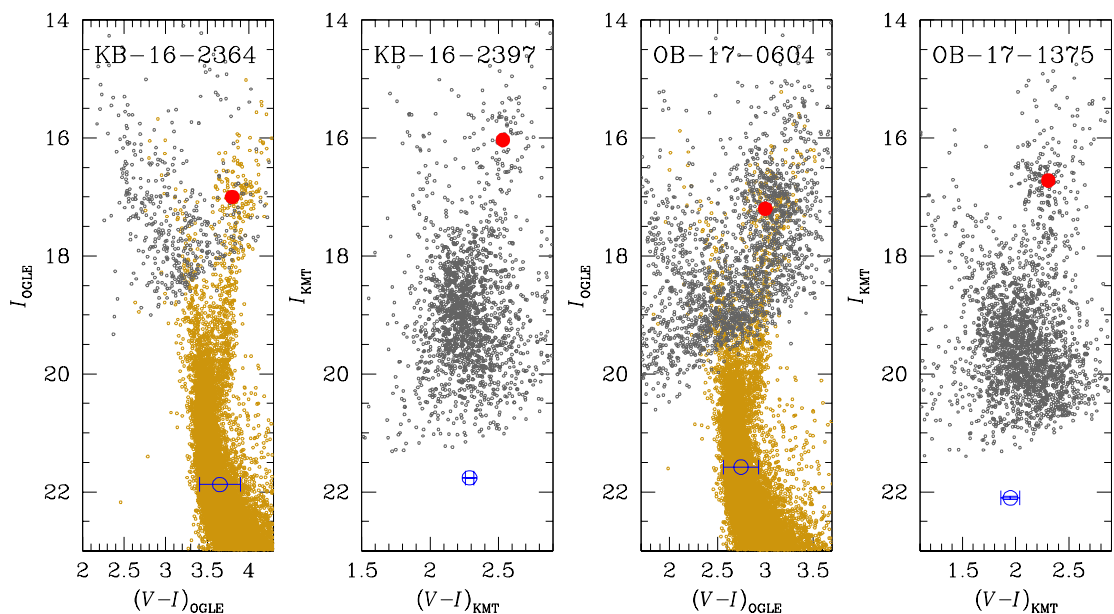}
\caption{
Positions of the source (marked by a blue empty dot with error bars) and red giant clump (RGC) 
centroid (red dot) in the instrumental color-magnitude diagrams.  For the events KMT-2016-BLG-2397 
and OGLE-2017-BLG-1375, the CMDs are constructed using the pyDIA photometry of the KMTC data set.  
For KMT-2016-BLG-2364 and OGLE-2017-BLG-0604, the CMDs are based on the combinations of the OGLE 
(grey points) and {\it HST} (yellow points) data sets.
}
\label{fig:eleven}
\end{figure*}

\begin{table*}[htb]
\small
\caption{Source color, magnitude, angular radius, Einstein radius, and proper motion\label{table:six}}
\begin{tabular}{lcccc}
\hline\hline
\multicolumn{1}{c}{Quantity}     &
\multicolumn{1}{c}{KMT-2016-BLG-2364}     &
\multicolumn{1}{c}{KMT-2016-BLG-2397}     &
\multicolumn{1}{c}{OGLE-2017-BLG-0604}    &
\multicolumn{1}{c}{OGLE-2017-BLG-1375}    \\
\hline
$(V-I, I)               $    &  $(3.65\pm 0.25, 21.88\pm 0.05) $ &  $(2.29\pm 0.03, 21.76\pm 0.01) $   &  $(2.75\pm 0.19, 21.58\pm 0.03) $  &  $(1.95\pm 0.09, 22.10\pm 0.02)$  \\
$(V-I, I)_{\rm RGC}     $    &  $(3.80, 17.01)                 $ &  $(2.53, 16.03)                 $   &  $(3.01, 17.20)                 $  &  $(2.31, 16.72)                $  \\
$(V-I, I)_{{\rm RGC},0} $    &  $(1.06, 14.39)                 $ &  $(1.06, 14.23)                 $   &  $(1.06, 14.48)                 $  &  $(1.060, 14.43)               $  \\
$(V-I, I)_0             $    &  $(0.91\pm 0.25, 19.26\pm 0.05) $ &  $(0.82\pm 0.03, 19.96\pm 0.01) $   &  $(0.81\pm 0.19, 18.86\pm 0.03) $  &  $(0.70\pm 0.09, 19.80\pm 0.02)$  \\
$\theta_*$ ($\mu$as)         &  $0.55\pm 0.14                  $ &  $0.36\pm 0.03                  $   &  $0.59\pm 0.12                  $  &  $0.34\pm 0.04                 $  \\
$\thetae$ (mas)              &  $> 0.18                        $ &  $> 0.24                        $   &  $0.98\pm 0.25                  $  &  $0.99\pm 0.13                 $  \\
$\mu$ (mas yr$^{-1}$)        &  $> 3.23                        $ &  $> 1.66                        $   &  $6.48\pm 1.65                  $  &  $3.56\pm 0.45                 $  \\
\hline
\end{tabular}
\end{table*}

Because the event timescale, $\te\gtrsim 100$~days, is considerably long, we check the feasibility 
of measuring the microlens parallax by conducting additional modeling of the light curve 
considering the microlens-parallax effect.  Because it is known that the microlens-parallax effect 
can be correlated with the effect of the lens orbital motion \citep{Batista2011, Skowron2011, Han2016}, 
we additionally consider the orbital motion of the lens in the modeling.  Considering the lens orbital 
motion requires to include two additional parameters $ds/dt$ and $d\alpha/dt$, which represent the 
change rates of the binary separation and the source trajectory angle, respectively.  We also check 
the ``ecliptic degeneracy'' between the pair of solutions with $u_0>0$ and $u_0<0$ caused by the 
mirror symmetry of the lens system configuration with respect to the binary axis \citep{Skowron2011}.  
The lensing parameters of the two solutions subject to this degeneracy are roughly 
related by
$(u_0, \alpha, \pien, d\alpha)\leftrightarrow -(u_0, \alpha, \pien, d\alpha)$.

Figure~\ref{fig:ten} shows the distributions of points in the MCMC chain on the $\piee$--$\pien$ 
parameter plane for the close (with $u_0<0$, left panel) and wide (with $u_0>0$, right panel) 
solutions.  We note that the corresponding solutions with opposite signs of $u_0$ exhibit similar 
distributions.  The distributions show that $\piee$ is relatively well constrained, but the uncertainty 
of $\pien$ is substantial, and this results in a large uncertainty of $\pie=(\pien^2+\piee^2)^{1/2}$.  
The uncertainties of the lens-orbital parameters, i.e., $ds/dt$ and $d\alpha/dt$, are also very large.  
As a result, the improvement of the fit by the higher-order effects is minor with $\Delta\chi^2< 8$ 
for all solutions.  In Table~\ref{table:five}, we list the lensing parameters estimated by considering 
the higher-order effects.  We note that the lens-orbital parameters are not listed in the table because 
they are poorly constrained.  It is found that the basic lensing parameters ($t_0$, $u_0$, $\te$, $s$, 
$q$, $\alpha$) vary little from those of the standard solution with the consideration of the 
higher-order effects.

\section{Source Stars and Einstein Radius}\label{sec:four}

We check the feasibility of measuring the angular Einstein radii of the events.  The measurement 
of $\thetae$ requires to characterize the source color, $V-I$, from 
which the angular source radius $\theta_*$ is estimated and the angular Einstein radius 
is determined by $\thetae=\theta_*/\rho$.  We are able to measure the source colors for 
KMT-2016-BLG-2397 and OGLE-2017-BLG-1375 using the usual method from the regression of 
$V$- and $I$-band magnitudes of data with the change of the lensing magnification 
\citep{Gould2010a}.  However, the source colors for KMT-2016-BLG-2364 and OGLE-2017-BLG-0604 
cannot be measured using this method, because the $V$-band magnitudes of the events cannot be 
securely measured due to the faintness of the source together with the severe $V$-band extinction 
of the fields, although the $I$-band magnitudes are measured.  For the latter two events, we 
estimate the source color using the {\it Hubble Space Telescope} ({\it HST}) color-magnitude 
diagram (CMD) \citep{Holtzman1998}.  In this method, the ground-based CMD is aligned with the 
{\it HST} CMD  using the red giant clump (RGC) centroids in the individual CMDs, and then we 
estimate the range of the source color as the width (standard deviation) of the main-sequence 
branch in the {\it HST} CMD for a given $I$-band brightness difference between the source and 
RGC centroid  \citep{Bennett2008, Shin2019}.  Besides the $V-I$ color, estimating $\thetae$ also 
requires measuring the normalized source radius $\rho$.  This is done for the events 
OGLE-2017-BLG-0604 and OGLE-2017-BLG-1375.  For the events KMT-2016-BLG-2364 and KMT-2016-BLG-2397, 
on the other hand, we can place only the upper limits on $\rho$, and thus set the lower limits 
of $\thetae$ for these two events.

For the $\theta_*$ estimation, we use the method of \citet{Yoo2004}.  According to this method, 
$\theta_*$ is estimated based on the extinction corrected (de-reddened) source color and magnitude,
 $(V-I, I)_0$, which are estimated using the RGC centroid, for which its de-reddened 
color and magnitude, $(V-I, I)_{{\rm RGC}0}$, are known, as a reference.  Following the procedure 
of the method, we first locate the source and RGC centroid in the instrumental CMD of stars in the 
vicinity of the source, measure the offsets in color, $\Delta (V-I)$, and brightness, $\Delta I$, 
of the source from the RGC centroid, and then estimate the de-reddened source color and magnitude by
\begin{equation}
(V-I, I)_0=(V-I, I)_{{\rm RGC},0} + \Delta (V-I, I).
\label{eq2}
\end{equation}
In this process, we use the reference values of $(V-I, I)_{{\rm RGC},0}$ estimated by 
\citet{Bensby2013} and \citet{Nataf2013}.

Figure~\ref{fig:eleven} shows the locations of the source and RGC centroid in the  
ground-based instrumental CMDs of stars (grey dots) around the source stars of the individual events.  
For the events KMT-2016-BLG-2364 and OGLE-2017-BLG-0604, for which the $V$-band source colors cannot 
be measured, we additionally present the {\it HST} CMDs (yellow dots), from which the source colors 
are derived.  In Table~\ref{table:six}, we list the values of the instrumental color and brightness 
for the source stars, $(V-I,I)$, and the RGC centroids, $(V-I,I)_{\rm RGC}$, and the dereddened 
color and magnitude of the source star, $(V-I,I)_0$, for the individual events.  The measured source 
colors and magnitudes indicate that the source stars have spectral types later than G2.

We estimate $\theta_*$ based on the measured source colors and magnitudes.  For this, we first 
convert $V-I$ into $V-K$ using the color-color relation of \citet{Bessell1988} and then estimate 
$\theta_*$ using the $(V-K)/\theta_*$ relation of \citet{Kervella2004}.  We determine $\thetae$ 
from the estimated $\theta_*$ by using the relation $\thetae=\theta_*/\rho$, and the relative 
lens-source proper motion is estimated from the combination of $\thetae$ and $\te$ by 
$\mu = \thetae/\te$.  We summarize the estimated values of $\theta_*$, $\thetae$, and $\mu$ in 
Table~\ref{table:six}.  We note that the measured $\thetae$ values for the events OGLE-2017-BLG-0604 
and OGLE-2017-BLG-1375 are about 2 times bigger than the value of a typical lensing event produced 
by a lens located roughly halfway between the source and lens, and this suggests that the lenses 
for these events are likely to be located close to the observer.

\begin{table}[t]
\small
\caption{
Availability of lensing observables
\label{table:seven}}
\begin{tabular*}{\columnwidth}{@{\extracolsep{\fill}}lccc}
\hline\hline
\multicolumn{1}{c}{Parameter}      &
\multicolumn{1}{c}{$\te$}          &
\multicolumn{1}{c}{$\thetae$}      &
\multicolumn{1}{c}{$\pie$}         \\
\hline
KMT-2016-BLG-2364      &  o   &   $\triangle$  &   x            \\
KMT-2016-BLG-2397      &  o   &   $\triangle$  &   x            \\
OGLE-2017-BLG-0604     &  o   &        o       &   x            \\
OGLE-2017-BLG-1375     &  o   &        o       &   $\triangle$  \\
\hline
\end{tabular*}
\tablefoot{
The notations ``o'' and ``x'' indicate that the observable is ``available''
and ``unavailable'', respectively. The notation ``$\triangle$'' implies that the observable is
measured but with a fairly large uncertainty.
}
\end{table}

\section{Physical Lens Parameters}\label{sec:five}

The lensing observables that can constrain the physical lens parameters of the mass and distance 
are $\te$, $\thetae$, and $\pie$.  The event timescale is rountinely measurable for most events,
but $\thetae$ and $\pie$ are measurable only for events satisfying specific conditions.  If all 
these observables are measured, the physical lens parameters are uniquely determined by
\begin{equation}
M={\thetae \over \kappa\pie};\qquad
D_{\rm L}= {{\rm au} \over \pie\thetae + \pi_{\rm S}},
\label{eq3}
\end{equation}
where $\kappa=4G/(c^2{\rm au})$, $\pi_{\rm S}={\rm au}/D_{\rm S}$, and $D_{\rm S}$ denotes the 
source distance.  For the analyzed events in this work, the availability of the observables 
varies depending on the events.  The event timescales are measured for all events.  The angular 
Einstein radii are measured for OGLE-2017-BLG-0604 and OGLE-2017-BLG-1375, but only the lower 
limits are set for KMT-2016-BLG-2364 and KMT-2016-BLG-2397.  The microlens parallax is measured 
only for OGLE-2017-BLG-1375, although the uncertainty of the measured $\pie$ is fairly large.  
We summarize the available observables for the individual events in Table~\ref{table:seven}.  
Considering the incompleteness of the information for the unique determinations of $M$ and $\dl$, 
we estimate the physical lens parameters by conducting Bayesian analyses with the constraint 
provided by the available observables of the individual events and using the priors of the lens 
mass function, physical, and dynamical Galactic models.

\begin{figure}
\centering
\includegraphics[width=\columnwidth]{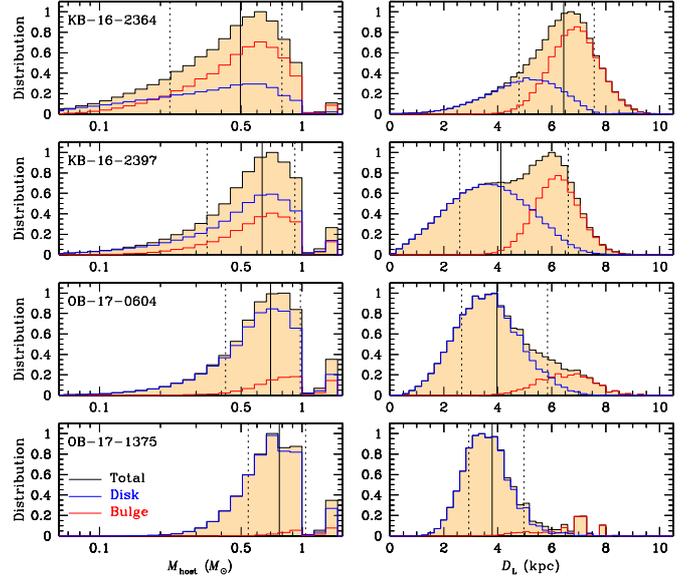}
\caption{
Posteriors of the host mass $M_{\rm host}$ (left panels)  and the distance $\dl$ (right panels) 
to the lens for the individual events obtained by Bayesian analyses.  In each panel, the blue 
and red distributions represent the contributions by the disk and bulge lens populations, 
respectively, and the black distribution is the sum of the contributions by two lens populations.  
The solid vertical line indicates the median value, and the two dotted vertical lines represent 
the $1\sigma$ range of the distribution.
}
\label{fig:twelve}
\end{figure}

\begin{table*}[thb]
\small
\caption{Physical lens parameters\label{table:eight}}
\begin{tabular}{llccccc}
\hline\hline
\multicolumn{3}{c}{Event}                       &
\multicolumn{1}{c}{$M_{\rm host}$ ($M_\odot$)}  &
\multicolumn{1}{c}{$M_{\rm p}$ ($M_{\rm J}$)}   &
\multicolumn{1}{c}{$\dl$(kpc)}                  &
\multicolumn{1}{c}{$a_\perp$ (au)}             \\
\hline
KMT-2016-BLG-2364      &                      &             &  $0.50^{+0.40}_{-0.27}$   &  $3.93^{+3.16}_{-2.17} $   &   $6.44^{+1.13}_{-1.60}$      &  $2.63^{+0.46}_{-0.65}$        \\
\hline                                                                                                                                                                                   
KMT-2016-BLG-2397      & Close                &             &  $0.62^{+0.29}_{-0.30}$   &  $2.42^{+1.12}_{-1.16} $   &   $5.09^{+1.50}_{-2.14}$      &  $2.83^{+0.84}_{-1.19}$        \\
                       & Wide                 &             &  $0.64^{+0.29}_{-0.30}$   &  $2.63^{+1.18}_{-1.23} $   &   $4.88^{+1.65}_{-2.18}$      &  $3.64^{+1.22}_{-1.61}$        \\
\hline                                                                                                                                                                                 
OGLE-2017-BLG-0604     &                      &             &  $0.70^{+0.28}_{-0.28}$   &  $0.51^{+0.21}_{-0.21} $   &   $3.95^{+1.87}_{-1.31}$      &  $4.06^{+1.91}_{-1.34}$        \\
\hline                                                                                                                                                                                 
OGLE-2017-BLG-1375     & Close                &  ($u_0>0$)  &  $0.77^{+0.27}_{-0.23}$   &  $10.33^{+3.61}_{-3.07}$   &   $3.79^{+1.18}_{-0.86}$      &  $2.97^{+0.92}_{-0.67}$        \\
                       &                      &  ($u_0<0$)  &  $0.81^{+0.63}_{-0.24}$   &  $11.28^{+8.73}_{-3.38}$   &   $3.93^{+1.49}_{-0.91}$      &  $3.04^{+1.16}_{-0.75}$        \\
                       & Wide                 &  ($u_0>0$)  &  $0.85^{+0.59}_{-0.23}$   &  $13.16^{+9.17}_{-3.62}$   &   $4.07^{+1.35}_{-0.86}$      &  $4.71^{+1.56}_{-1.00}$        \\
                       &                      &  ($u_0<0$)  &  $0.85^{+0.60}_{-0.23}$   &  $13.27^{+9.39}_{-3.58}$   &   $4.08^{+1.42}_{-0.87}$      &  $4.72^{+1.64}_{-1.00}$        \\
\hline                                                                                                                                                                    
\end{tabular}
\end{table*}

A Bayesian analysis is carried out by producing a large number ($4\times 10^7$) of artificial lensing 
events from a Monte Carlo simulation using the priors.  The priors of the Galactic model are based on 
the modified version of the \citet{Han2003} model for the physical matter density distribution, and 
the \citet{Han1995} model for the distribution of the relative lens-source transverse speed. For the 
mass function, we use the \citet{Zhang2020} model for stellar and brown-dwarf lenses and the 
\citet{Gould2000} model for remnant lenses, i.e., white dwarfs, neutron stars, and black holes.  
For more details of the models, see section~5 of \citet{Han2020}. With the produced events, we then 
obtain the posteriors for $M$ and $\dl$ by constructing the probability distributions of events by 
applying the constraints available for the individual events.  Although $\thetae$ values are not 
uniquely measured for KMT-2016-BLG-2364 and KMT-2016-BLG-2397, we apply the constraint of their lower 
values.  For OGLE-2017-BLG-1375, we apply a two-dimensional constraint of $\pivec_{\rm E}$, i.e., 
$(\pien, \piee)$.  With the constructed probability distributions, we then choose representative 
values of the physical parameters as the median of the distributions and estimate the uncertainties 
of the parameters as the 16\% and 84\% ranges of the distributions.

In Figure~\ref{fig:twelve}, we present the posteriors of the host mass (left panels), $M_{\rm host}$, 
and the distance to the lens (right panels) for the individual events.  In each panel, the blue and red 
distributions represent the contributions by the disk and bulge lens populations, respectively, and the 
black distribution is the sum of the contributions by two lens populations. The solid vertical line 
indicates the median value, and the two dotted vertical lines represent the $1\sigma$ range of the 
distribution.

The estimated masses of the lens components ($M_{\rm host}$ and $M_{\rm p}=qM_{\rm host}$), distance 
($\dl$), and projected planet-host separation ($a_\perp=s\dl \thetae$) are summarized in 
Table~\ref{table:eight}.  The masses of the hosts and planets are in the ranges 
$0.50\lesssim M_{\rm host}/M_\odot\lesssim 0.85$ and $0.5 \lesssim M_{\rm p}/M_{\rm J}\lesssim 13$, 
respectively, indicating that all planetary systems are composed of giant planets and host stars with 
subsolar masses.  To be noted is that the lower mass component of OGLE-2017-BLG-1375L lies at around 
the boundary between planets and brown dwarfs, i.e., $\sim 13~M_{\rm J}$ \citep{Boss2007}.  The 
lenses are located in the distance range of $3.8 \lesssim \dl/{\rm kpc}\lesssim 6.4$.  We note that 
the lenses of OGLE-2017-BLG-0604 and OGLE-2017-BLG-1375 (both with $\dl\sim 4~{\rm kpc}$) are likely 
to be in the Galactic disk.

We note three of the four planetary hosts analyzed in this paper can almost certainly be resolved 
by adaptive optics (AO) observations on next-generation ``30m'' telescopes at AO first light (roughly 
2030). That is, according to Figure~\ref{fig:twelve}, each host of the four lens systems has only a 
tiny probability of being non-luminous. And, with exception of KMT-2016-BLG-2397, all lensing events 
have relative lens-source proper motions $\mu\gtrsim 3~{\rm mas}~{\rm yr}^{-1}$. Hence, in 2030, they 
will be separated from the source by $\Delta\theta\gtrsim 40$~mas. It is very likely that KMT-2016-BLG-2397 
can also be resolved, unless it is extremely close to the limit that we report in Table~\ref{table:six}. 
Given that $t_{\rm E}$ is well measured for all four events, such a proper-motion measurement will 
immediately yield a $\theta_{\rm E}$ measurement for the two events that do not already have one, and 
a more precise $\theta_{\rm E}$ measurement for the other two. Combined with the $K$-band source flux 
measurement from the AO observations themselves, this will yield good estimates of the lens mass and 
distance. In the case of OGLE-2017-BLG-1375, the one-dimensional parallax measurement will enable 
even more precise determinations \citep{Gould2014}.

\section{Summary and Conclusion}\label{sec:six}

For a solid demographic census of microlensing planetary systems based on more complete sample, we 
investigated microlensing data in the 2016 and 2017 seasons obtained by the KMTNet and OGLE surveys 
to search for missing or unpublished planetary microlensing events.  From this investigation, we found 
four planetary events: KMT-2016-BLG-2364, KMT-2016-BLG-2397, OGLE-2017-BLG-0604, and 
OGLE-2017-BLG-1375.  It was found that the events share a common characteristic that the sources were 
faint stars.  We presented the detailed procedure of modeling the observed light curves conducted to 
determine lensing parameters and presented models for the individual lensing events.  We then carried 
out Bayesian analyses for the individual events using the available observables that could constrain 
the physical lens parameters of the mass and distance.  From these analyses, it was found that the 
masses of the hosts and planets were in the ranges $0.50\lesssim M_{\rm host}/M_\odot\lesssim 0.85$ 
and $0.5 \lesssim M_{\rm p}/M_{\rm J}\lesssim 13$ respectively, indicating that all planets were giant 
planets around host stars with subsolar masses.  It was estimated that the distances to the lenses were 
in the range of $3.8 \lesssim \dl/{\rm kpc}\lesssim 6.4$.  It was found that the lenses of OGLE-2017-BLG-0604 
and OGLE-2017-BLG-1375 were likely to be in the Galactic disk.

\begin{acknowledgements}
Work by CH was supported by the grants  of National Research Foundation of Korea 
(2017R1A4A1015178 and 2020R1A4A2002885).
Work by AG was supported by JPL grant 1500811.
This research has made use of the KMTNet system operated by the Korea
Astronomy and Space Science Institute (KASI) and the data were obtained at
three host sites of CTIO in Chile, SAAO in South Africa, and SSO in
Australia.
The OGLE project has received funding from the National Science Centre, Poland, grant
MAESTRO 2014/14/A/ST9/00121 to AU.
\end{acknowledgements}

%
%

\end{document}